\documentclass[conference]{IEEEtran}
\IEEEoverridecommandlockouts

\usepackage{booktabs}

\usepackage{url}
\usepackage{subfig}
\usepackage{graphicx}
\graphicspath{{figures/}}

\usepackage{todonotes}

\usepackage[noadjust]{cite}
\usepackage{amsmath,amssymb,amsfonts}
\usepackage{algorithmic}
\usepackage{graphicx}
\usepackage{textcomp}
\usepackage{makecell}
\usepackage{adjustbox}
\usepackage{balance}  
\usepackage{xcolor}
\usepackage{multirow}
\def\BibTeX{{\rm B\kern-.05em{\sc i\kern-.025em b}\kern-.08em
    T\kern-.1667em\lower.7ex\hbox{E}\kern-.125emX}}
\begin{document}

\title{Mining Student Responses to Infer Student Satisfaction Predictors}


\author{\IEEEauthorblockN{Farzana Afrin, Mohammad Saiedur Rahaman, Margaret Hamilton
}
\IEEEauthorblockA{Computer Science and Information Technology, School of Science, RMIT University, VIC, Australia
}

Email: \{farzana.afrin, saiedur.rahaman, margaret.hamilton\}@rmit.edu.au
}

\maketitle

\begin{abstract}
The identification and analysis of student satisfaction is a challenging issue. This is becoming increasingly important since a measure of student satisfaction is taken as an indication of how well a course has been taught. However, it remains a challenging problem as student satisfaction has various aspects. In this paper, we formulate the student satisfaction estimation as a prediction problem where we predict different levels of student satisfaction and infer the influential predictors related to course and instructor.  We present five different aspects of student satisfaction in terms of 1) course content, 2) class participation, 3) achievement of initial expectations about the course, 4) relevancy towards professional development, and 5) if the course connects them and helps to explore the real-world situations. We employ state-of-the-art machine learning techniques to predict each of these aspects of student satisfaction levels. For our experiment, we utilize a large student evaluation dataset which includes student perception using different attributes related to courses and the instructors. Our experimental results and comprehensive analysis reveal that student satisfaction is more influenced by course attributes in comparison to instructor related attributes.

\end{abstract}


\section{Introduction}

Student satisfaction is regarded as a vital component for measuring the success of education service providers such as universities and post-secondary institutions. Proving that higher education (HE) provides a good experience for students is important for promoting and advertising the University both in the home country and internationally. It is considered important to measure the efficiency, quality, impact of the teaching provided by evaluating its effect on students. Student satisfaction ratings play an important role in identifying main areas of strength and weakness for further improvement by the HE service providers. Eventually, it can help them to better meet student expectations by identifying areas of student interests and needs. The HE providers are required to understand the key elements or factors that constitute student satisfaction,  not only around the overall program experience but student satisfaction about their individual course units as well. 

Since measuring the overall satisfaction cannot provide an accurate estimation of individual student satisfaction, all instructors are encouraged to survey their students at the end of every course. The collected data can be utilized to enhance the quality of academic institutions \cite{Klemen2015}. The two surveys,  the Student Experience Survey (SES) and the Course Experience Survey (CES) are increasingly being used to measure teaching and university performance in the provision of HE. However, the data collected from student experience and engagement surveys raise questions in terms of quality, reliability, and validity, and may not be used as a piece of concrete evidence in decision-making for higher education institutions \cite{Klem2015}. A recent research investigation has identified the incorporation of many irrelevant questions in the course evaluation surveys which influence the responses and reduce the quality of collected data to be used as an indicator of desirable performance measure \cite{7469785}. Therefore, it is important to identify and analyze relevant factors for informed decision making.

The use of various data-driven solutions are becoming increasingly popular for understanding educational and administrative problems in higher education \cite{5524021,7469785}.
The aim of educational data mining (EDM) is to discover valuable hidden insights from data related to educational services which may be difficult and very time consuming if done manually \cite{MOHAMAD2013320}. The use of EDM can lead towards more benefits and impacts for students, educators and education service providers by upgrading the levels of academic achievement and success for students in a more effective and efficient manner \cite{SHAHIRI2015414}. 

In this paper, we utilize state-of-the-art machine learning techniques to investigate five aspects of student satisfaction around course evaluation, class participation, course expectation, course relevancy and professional development (as it relates to real-world employment). Specifically, we build models using student evaluation data to predict each of these satisfaction indicators. We further investigate the crucial attributes associated with each aspect of student satisfaction. Since the satisfaction is very challenging to define and there might be various other factors that influence student satisfaction, we limit our investigation within the course and instructor related predictors. The main research question we address in this paper is as below: \\

\noindent
\textit{``What are the predictors related to instructor and course that influence different aspects of student satisfaction?"}\\

The rest of this paper is organized as follows: Section 2 provides a review of related literature. Section 3 provides a formal definition to the research problem. Section 4 describes the analytic approach used for our experiment and analysis. In Section 5, we show experimental results. Finally, Section 6 concludes the paper.


\section{Review of Related Literature}

The wide use of EDM is evident from the increasing development of methods for exploring large-scale data generated within the higher education sector. A comprehensive survey of the key components of EDM has been presented in \cite{PENAAYALA20141432}. The research identified three categories: tasks, methods, and algorithms as three core building blocks of EDM related research. The use of data mining and machine learning is well studied in educational technology research as well where the aim is to understand how students interact with technology \cite{ANGELI2017226}. The researchers can highlight the existence of several distinct patterns in students' learning experiences.


To analyze the performance of undergraduate students, different data mining techniques have been used in \cite{ASIF2017177}. The study predicted the final academic achievements of the students at the end of their four-year bachelor degree. These authors also analyzed the progression of students throughout their academic years. Another study used predictive analysis to infer the academic performance of students \cite{FERNANDES2018}. Yet another research paper applied data mining techniques on historical student course grade data to model student performance \cite{BURGOS2018541}, demonstrating the determination of course drop out of a student. A tutoring action plan is proposed based on the modeling for the prevention of academic dropouts. Different classification techniques are evaluated to predict the slow learners in an educational setup \cite{KAUR2015500}. The effectiveness of using machine learning techniques to predict the performance of students has been discussed in \cite{7684167}. A atudent evaluation dataset has been leveraged to estimate the performance of the course instructor in \cite{AHMED2016137}. This research uses various machine learning techniques including J48, Multilayer Perception, Na\"{i}ve Bayes, and Sequential Minimal Optimization. Another research paper that investigates the performance of instructors is presented by leveraging the data from course evaluation questionnaire and applying data mining algorithms in \cite{7469785}. This research mainly considers the perception of students and found that there are many irrelevant questions in the course evaluation survey which cannot be related to the instructors' performance. Another research study utilizes clustering and sequential pattern mining techniques to analyze the careers of university graduates by \cite{CAMPAGNI20155508}. This research is based on a real case study which introduces the concept an ideal career to represent the career of a model student. Another review to predict student performance using several machine learning techniques is presented in \cite{SHAHIRI2015414}.

Several research projects have been conducted on various factors to measure student satisfaction after the course has finished. Student satisfaction in a typical computer science course has been researched in
\cite{DBLP:journals/corr/AfrinRRR15}. This research identifies the proper course planning, student encouragement, satisfactory coverage and relevance of the course to the real-world software development scenario as the four key factors of student satisfaction. To understand general measures of student satisfaction from student-opinion data, the regression and decision tree models are utilized in \cite{Thomas2004}. This research identifies distinct predictors for each measure which cannot be used interchangeably. The social integration is found to have an immense effect on student satisfaction for those students who are not highly engaged academically. Another comprehensive use of data mining techniques for gaining insight into student satisfaction is presented in \cite{DEJAEGER2012548}. This research highlights the fact that data mining techniques can be useful to identify and act on a small number of predictors that require attention in order to increase and manage student satisfaction.

The student satisfaction research discussed above considers mainly one aspect of satisfaction. This paper infers and provides an analysis of five different aspects of student satisfaction by applying educational data mining techniques. We aim to investigate course and instructor related factors to identify the ones with the largest impact on student satisfaction.

\section{Research Problem Formulation}
We have two main goals in this research. First, we investigate whether different aspects of student satisfaction can be predicted using various combinations of features related to course and instructor. To achieve our goal, we train a set of machine learning algorithms using a portion of the historical student evaluation data and log the prediction performance for unknown instances. Second, we infer and analyze influential predictors for student satisfaction. We formalize our research problems as below:  

\begin{itemize}
    \item Let, $S_{i} = \{1, 2, 3, 4, 5\}$ be the set of five satisfaction levels of satisfaction aspect $S_{i}$, where 1 indicates lowest satisfaction and 5 indicates highest satisfaction. In the student evaluation dataset, each instance $x$ is described by a $d$-dimensional vector of attributes $R^{d}$ related to course and instructor, and one satisfaction level. If $f(.)$ is the satisfaction level prediction function for an unknown instance of $d$-features, $f(.)$ predicts $\hat{S}(x_{q})$ such as $f(x_{q}):R^{d} \rightarrow \hat{S}(x_{q})$  where $\hat{S}(x_{q})$ is the predicted satisfaction level of the query instance $x_{q}$. 
    
    \item Given a set of predictors related to course and instructor, we can identify the factors from the student evaluation dataset that influence the prediction outcomes. We further investigate the extent to which these factors, which are either course or instructor related, drive the student satisfaction. 
\end{itemize}

\section{Analytic Approach}

The analytic approach used in this paper has three core components that are interrelated with each other: 1) Investigation of a student evaluation dataset, 2) Definition and prediction of student satisfaction utilizing the student evaluation dataset, and 3) Inference of influential predictors for student satisfaction. 

\subsection{The Student Evaluation Data}

\begin{table*}[h!]
\centering
\caption{Description of Fields in the Student Evaluation Dataset}
\label{data-table}
\renewcommand*{\arraystretch}{1.1}
\small
\begin{tabular}{p{1.5cm}p{13cm}p{1.5cm}}
\toprule
Fields  & Description Description of Fields                               &  Values            \\ \midrule
instr      & Instructor identifier                                       & \{1,2,3\}          \\
class      & Course code descriptor                                      & \{1-13\}           \\
repeat     & Number of times the student has taken this course             & \{0,1,2,3\}    \\
attendance & Code of the level of attendance                               & \{0,1,2,3,4\}  \\
difficulty & Level of difficulty of the course as perceived by the student & \{1,2,3,4,5\}      \\ 
Q1 & The semester course content, teaching method and evaluation system were provided at the start & \{1,2,3,4,5\}\\
Q2 & The course aims and objectives were clearly stated at the beginning.& \{1,2,3,4,5\}\\
Q3 & The course was worth the amount of credit assigned to it. & \{1,2,3,4,5\}\\
Q4 & The course was taught according to the announced syllabus. & \{1,2,3,4,5\}\\
Q5 &	The class discussions, homework assignments, applications and studies were satisfactory. & \{1,2,3,4,5\}\\
Q6 & The textbook and other courses resources were sufficient and up to date.	& \{1,2,3,4,5\}\\
Q7 & The course allowed field work, applications, laboratory, discussion and other studies. & \{1,2,3,4,5\}\\
Q8 & The quizzes, assignments, projects and exams contributed to the learning.	& \{1,2,3,4,5\}\\
Q9 & I greatly enjoyed the class and was eager to actively participate. & \{1,2,3,4,5\}\\
Q10 & Initial expectations about the course were met at the end. & \{1,2,3,4,5\}\\
Q11 & The course was relevant and beneficial to my professional development. & \{1,2,3,4,5\}\\
Q12 & The course helped me look at life and the world with a new perspective. & \{1,2,3,4,5\}\\
Q13 & Instructor's knowledge was relevant and up to date. & \{1,2,3,4,5\}\\
Q14 & Instructor came prepared for classes. & \{1,2,3,4,5\}\\
Q15 & Instructor taught in accordance with the announced lesson plan. & \{1,2,3,4,5\}\\
Q16 & Instructor was committed to the course and was understandable. & \{1,2,3,4,5\}\\
Q17 & Instructor arrived on time for classes. & \{1,2,3,4,5\}\\
Q18 & Instructor has a smooth and easy to follow delivery/speech. & \{1,2,3,4,5\}\\
Q19 & Instructor made effective use of class hours. & \{1,2,3,4,5\}\\
Q20 & Instructor explained the course and was eager to be helpful to students. & \{1,2,3,4,5\}\\
Q21 & Instructor demonstrated a positive approach to students. & \{1,2,3,4,5\}\\
Q22 & Instructor was open and respectful of the views of students. & \{1,2,3,4,5\}\\
Q23 & Instructor encouraged participation in the course. & \{1,2,3,4,5\}\\
Q24 & Instructor gave relevant homework assignments/projects, and helped/guided students. & \{1,2,3,4,5\}\\
Q25 & Instructor responded to questions about the course inside and outside. & \{1,2,3,4,5\}\\
Q26 & Instructor's evaluation (exams, projects, assignments) effectively measured the course objectives. & \{1,2,3,4,5\}\\
Q27 & Instructor provided solutions to exams and discussed them with students. & \{1,2,3,4,5\}\\
Q28 & Instructor treated all students in a right and objective manner. & \{1,2,3,4,5\}\\
\bottomrule
\end{tabular}
\end{table*}

For our experiments and evaluation, we utilize a student evaluation dataset which is publicly available from the UCI repository and was contributed by the students of Gazi University in Ankara, Turkey \cite{GunduzFokoue:2013}. This dataset contains a total of 5820 evaluation scores provided by the students in response to 28 specific questions (Q1 to Q28) related to course and instructor. All the responses to these 28 questions are logged in a 5-point Likert's scale where 1 indicates lowest and 5 indicates highest. The dataset also contains student responses in regards to an additional 5 attributes: instructor code, course code, number of repeats (of this course), level of attendance and perceived difficulty of the course. The response to the difficulty of the course was given between 1 and 5. All these responses were captured for a total of 13 courses. The response to the number of repeats is a number between 0 and 4. The description of different fields of the dataset is given in Table \ref{data-table}. A comprehensive analysis of this student evaluation dataset is provided at the beginning of the experiment and discussion section.

\subsection{Define and Predict Student Satisfaction}
We consider five aspects of student satisfaction. After careful investigation, we decode the meanings of different questions in the student evaluation dataset. We consider that Q8, Q9, Q10, Q11 and Q12 (see Table \ref{data-table}) are the representative questions that capture five aspects of student satisfaction in terms of \textit{course content}, \textit{class participation}, \textit{achievement of initial expectations about the course}, \textit{relevancy towards professional development}, and \textit{the extent to which the course connects and helps the student explore real-world situations}. To predict these five aspects of student satisfaction, we train a set of machine learning techniques including Support Vector Machine (SVM), Multilayer Perceptron (MLP), Decision Tree, Random Forest, Decision Table, $k$-Nearest Neighbour ($k$-NN). Further details will be explained in the Experiment and Discussion section.

\subsection{Calculating Influential Predictors}
\label{influential-predictors}

There are several techniques applied to various domains to compute important predictors for informed decision making \cite{al2012construction,abdullah2012stock,rahaman2017queue,al2012knowledge, rahaman2018wait}. We identify the most influential predictors of student satisfaction using an information theoretic measure known as `Mutual Information' which measures the mutual dependence between two variables. In our case, one of the two variables is a feature while the other variable is the target (i.e. satisfaction level). We consider one aspect of satisfaction at a time during the calculation of mutual information between a prediction and the target. The higher the value of computed mutual information the better the influence of the predictor considered for computation. The mutual information between a predictor and the satisfaction level is calculated as below:

\begin{equation}
I(a_{j};S_i)=-\sum_{j,i} p(a_{j},S_i) \log \frac{p(a_{j},S_i)}{p(a_{j})p(S_i)}
\end{equation}

Here, $a_{j}$ is $j^{th}$ predictor of $i^{th}$ satisfaction aspect and $Si$ is any satisfaction aspect $i$; $p(a_{j},S_i)$ is the joint probability.


\section{Experiment and Discussion}

\subsection{Preliminary Analysis}

\begin{figure*}[t]
\centering
\includegraphics[width=6in, height=3in]{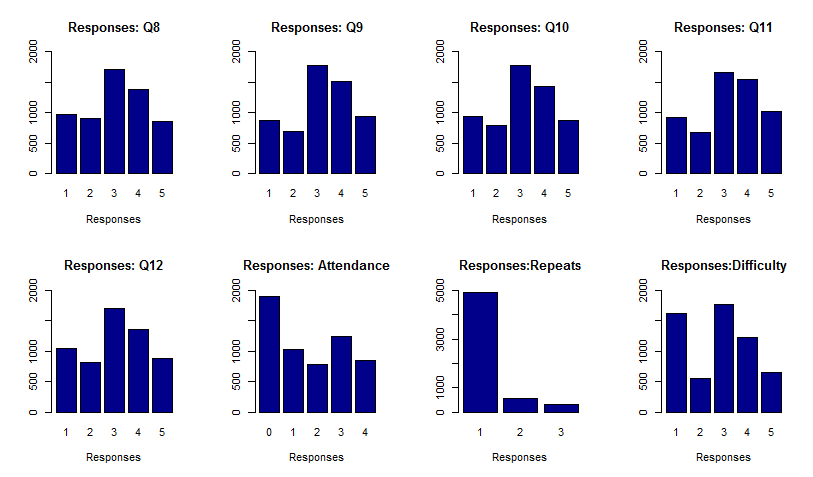}
 \caption{Distribution of response frequencies for five satisfaction aspects (i.e. Q8, Q9, Q10, Q11, Q12) along with attendance levels, number of repeats and perceived difficulty levels associated with different courses.}
\label{fig:distribution:all-responses}
\end{figure*}

At the beginning of our experiment with student evaluation data, we analyze the distribution of responses logged in the dataset against the five different aspects of student satisfaction. Figure \ref{fig:distribution:all-responses} illustrates the distribution of response frequencies for five satisfaction aspects (i.e. Q8, Q9, Q10, Q11, Q12) along with attendance levels, number of repeats and perceived difficulty levels associated with different courses. It should be noted that a response `5' in a satisfaction aspect indicates the frequency of respondents who are `very satisfied' while a response `1' indicates very `unsatisfied'. A response of `3' indicates the neutral views of the students. We can see an imbalance in the response frequencies for five satisfaction aspects i.e. response `3' is the most dominant response selected by the participant students followed by the response `4'. This issue needs to be considered while training machine learning algorithms since the training performance can be compromised by the label imbalanced problem if present in the dataset. The following subsection explains how we address this challenge.

We also can see from Figure \ref{fig:distribution:all-responses} that a large number of students had different attendance ratings as indicated by the attendance codes between 0 and 4. The attendance code 0 contributes most of the frequencies compared to others. Figure \ref{fig:distribution:all-responses} also highlights that the students mostly undertook the course only once. There are some students who repeated the course for two or three times. However, there are no student participants in this dataset who did not complete this course.

Students also provided their perceptions of the difficulty level of their specific course. As can be seen from Figure \ref{fig:distribution:all-responses}, a large number of students are neutral in their responses in terms of the difficulty level of the course which is represented by a difficulty rating of `3', followed by a large proportion who found it easy represented by `1' (i.e. least difficult). Also, there are a large number of students who found the course 'difficult' or `very difficult' as can be seen from the frequencies of responses `4' and `5'.

\subsection{Student Satisfaction Prediction}

In our experiments, we use the Weka \cite{hall09:_weka_data_minin_softw} toolkit for a set of classifiers including Support Vector Machine (SVM), Multilayer Perceptron (MLP), Decision Tree, Random Forest, Decision Table, and $k$-Nearest Neighbor ($k$-NN). In our Decision Tree implementation, we used information gain measure for top-down induction. An ensemble of 100 trees was used for the Random Forest. Default parameters were considered for the implementation of SVM, MLP and Decision Table. We used $k=5, 10$ for our implementation of the $k$-NN classifier.

To validate the performance of our satisfaction prediction, we use a stratified cross-validation. The reason of using stratified cross-validation is to ensure the presence of an equal number of target (i.e. satisfaction) labels in the training phase of each cross-validation. In other words, this is to reduce the effect of label imbalance in our data. For student satisfaction prediction, we conduct three sets of experiments as follows:

\begin{itemize}
    \item \textit{Predicting satisfaction using predictors related to GTS (Good Teaching Score) only}: GTS is one of the widely used metrics in many countries such as Australia to measure student satisfaction \cite{gts}. A GTS score consists of 6 main features that represent 6 key teaching areas. The responses for calculating the GTS score are collected by asking students to rate the following 6 statements: GTS-1) The teaching staff are extremely good at explaining things; GTS-2) The teaching staff normally give me helpful feedback on how I am going; GTS-3) The teaching staff motivate me to do my best work; GTS-4) The teaching staff work hard to make the course interesting; GTS-5) The teaching staff make a real effort to understand difficulties I might be having with my work; and GTS-6) The staff put a lot of time into commenting on my work. The GTS scores have been shown to correlate well with student satisfaction \cite{gts1}. 
    
    Since the student evaluation dataset used in this paper does not contain all of these 6 questions directly, we consider a total of 9 features that together can be used to describe the actual 6 GTS features. We consider that GTS-1 is equivalent to Q16 and Q18; GTS-2 is equivalent Q25; GTS-3 is equivalent Q28; GTS-4 is equivalent Q22 and Q23; GTS-5 is equivalent Q21; and GTS-6 is equivalent Q24 and Q27.
    
    The outcome of the student satisfaction prediction using GTS features only is given in Table \ref{table:gtsaccuracy}. Here the maximum prediction accuracies produced by the SVM algorithm for the satisfaction aspect features of Q8, Q10, Q11, and Q12 are 76.01\%, 77.85\%, 76.10\% and 75.34\% respectively. For the overall student satisfaction aspect represented by Q9, $k$-NN ($k$ =10) produces the highest prediction accuracy of 75.94\%.
        
    
    \item \textit{Predicting satisfaction using predictors related to instructor only}: In this set of experiments, we consider only those predictors related to instructors. These features are from Q13 to Q28 in the student evaluation dataset. The outcome of the student satisfaction prediction using instructor features only is given in Table \ref{table:q13-q28}. We can see that the maximum prediction accuracy is produced by the SVM algorithm for satisfaction aspect Q8 and is 76.98\%. For the overall student satisfaction aspect Q9, and Q11 $k$-NN ($k$ =10) produces highest prediction accuracies of 75.96\% and 76.89\% respectively. The highest accuracy for Q10 is 78.40\% which is given by the Decision Table algorithm while $k$-NN ($k$ =5) produces the highest prediction accuracy for Q12 (76.03\%).
    
    \item \textit{Predicting satisfaction using predictors related to both course and instructor}: In these experiments, we consider predictors related to the course (Q1 to Q7) and instructors (Q13 to Q28) together. We can see from table \ref{table:accuracy} that the maximum prediction accuracies produced by the SVM algorithm for the satisfaction aspects Q8, Q10, Q11, and Q12 are 84.16\%, 84.02\%, 80.82\% and 79.86\% respectively. For Q9, the Random Forest algorithm produces the highest accuracy of 80.86\%. 
\end{itemize}

\begin{table}[h]
\centering
\caption{Accuracy using Good Teaching Score (GTS) features only}
\label{table:gtsaccuracy}
\renewcommand*{\arraystretch}{1}
\begin{tabular}{llllll}
\toprule
               & Q8    & Q9     & Q10   & Q11   & Q12   \\\midrule
SVM            & \textbf{76.01} & 75.58  & \textbf{77.85} & \textbf{76.10} & \textbf{75.34} \\
MLP            & 73.37 & 72.81 & 76.0 & 73.12 & 73.88 \\
Decision Tree  & 75.24  & 74.52  & 76.82 & 74.97 & 74.78 \\
Random Forest  & 74.91 & 73.42  & 76.77 & 74.28 &  74.07\\
Decision Table & 75.77 & 75.26  & 77.34 & 75.09 & 74.74\\
k-NN (k = 5)   & 75.48 & 75.15  & 77.37 & 75.33 &  74.62\\
k-NN (k = 10)  & 75.96 & \textbf{75.94}  & 77.66 & 76.03 &  75.0\\\bottomrule
\end{tabular}
\end{table}

\begin{table}[h]
\centering
\caption{Accuracy using predictors related to instructors only (Q13-Q28)}
\label{table:q13-q28}
\renewcommand*{\arraystretch}{1}
\begin{tabular}{llllll}
\toprule
               & Q8    & Q9     & Q10   & Q11   & Q12   \\\midrule
SVM            & \textbf{76.98} & 75.89  & 78.37 & 76.54 & 75.96 \\
MLP            & 73.52 & 71.85  & 76.15 & 73.77 & 73.17 \\
Decision Tree  & 75.26 & 74.12  & 77.64 & 75.38 & 74.91 \\
Random Forest  & 75.67 & 74.47  & 77.73 & 76.17 & 75.17 \\
Decision Table & 75.74 & 74.71  & \textbf{78.40} & 75.96 & 75.93\\
k-NN (k = 5)   & 76.27 & 75.57  & 77.92 & 76.36 & \textbf{76.03} \\
k-NN (k = 10)  & 76.39 & \textbf{75.96}  & 78.20 & \textbf{76.89} & 75.84 \\\bottomrule
\end{tabular}
\end{table}

\begin{table}[h]
\centering
\caption{Accuracy of student satisfaction prediction using both course and instructor predictors. Note: GTS features are a subset of instructor related features.}
\label{table:accuracy}
\renewcommand*{\arraystretch}{1}
\begin{tabular}{llllll}
\toprule
               & Q8    & Q9     & Q10   & Q11   & Q12   \\\midrule
SVM            & \textbf{84.16} & 80.69  & \textbf{84.02} & \textbf{80.82} & \textbf{79.86} \\
MLP            & 81.51 & 78.281 & 81.46 & 77.75 & 77.23 \\
Decision Tree  & 82.0  & 78.83  & 81.05 & 78.61 & 77.97 \\
Random Forest  & 84.11 & \textbf{80.86}  & 83.76 & 80.67 & 79.07 \\
Decision Table & 82.32 & 79.85  & 82.65 & 79.79 & 78.45\\
k-NN (k = 5)   & 82.73 & 80.36  & 83.33 & 80.67 & 79.19 \\
k-NN (k = 10)  & 82.44 & 80.45  & 83.13 & 80.27 & 79.38 \\\bottomrule
\end{tabular}
\end{table}

In summary, we see from Tables \ref{table:gtsaccuracy}, \ref{table:q13-q28} and \ref{table:accuracy} that the combination of course and instructor related predictors produces the highest prediction accuracies for student satisfaction prediction. In other words, an improvement of prediction accuracies can be seen by considering the course and instructor related predictors in comparison to results using only GTS features or only instructor related features. This improvement is approximately 5\%-7\% across all the aspects of student satisfaction. This indicates that the course related features have more influence in predicting student satisfaction. We further investigate the most influential predictors of student satisfaction in the following subsection.


\subsection{Inferring Influential Predictors}

\begin{figure*}[t]
\centering
\includegraphics[width=6.2in]{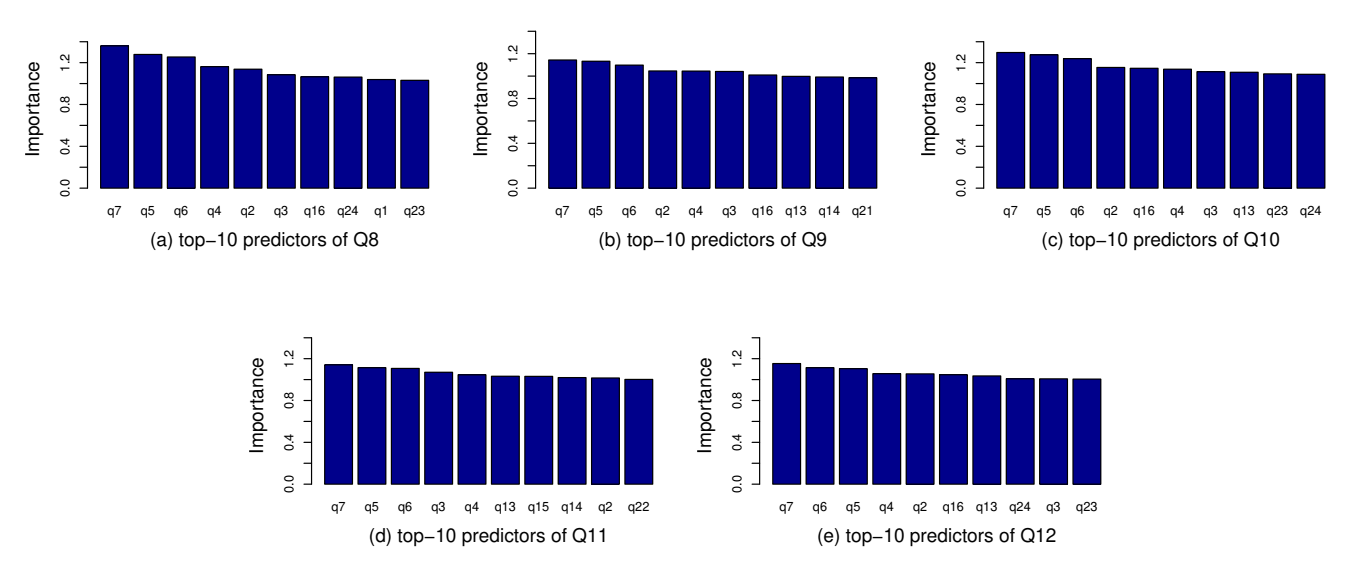}
 \caption{Important features for predicting satisfaction aspects (Q8 - Q12)}
\label{fig:important:all}
\end{figure*}

As explained in Section \ref{influential-predictors}, we calculate the mutual information between two variables (i.e. a predictor variable and the satisfaction label) to estimate the influence of any predictor on different aspects of student satisfaction. In this section, we list the top-10 influential predictors related to course (i.e. Q1 to Q7) and instructors (i.e. Q13 to Q28). It should be noted that the corresponding importance scores (i.e. mutual information scores) for all predictors were calculated by building a Random Forest classifier each time. To identify the top-10 predictors, we sorted the importance scores in descending order. We find that the top-10 predictors are a subset of a total of 14 predictors which includes Q1 to Q7, Q13 to Q16, and Q22 to 24. We plot the results in Figures \ref{fig:important:all} (a)-(e) to show the top-10 predictors of Q8, Q9, Q10, Q11, and Q12. There is only one occasion when one predictor related to instructor was found in the top-5 while the other four relate to the course.



 Figure \ref{fig:important:all} (a) shows that there are only four predictors related to instructor in the top-10 list that influence student satisfaction in terms of course content (Q8). All of the top-5 predictors are related to the course itself. Figure \ref{fig:important:all} (b) shows that student satisfaction in terms of class participation (Q9) is influenced by four predictors related to instructor as found in the list of top-10 predictors. Once again, as for Q8, all of the top-5 influential predictors are related to the course itself. There is only one instructor related predictor in the top-5 for predicting student satisfaction in terms of the achievement of initial expectations about the course (Q10) while the other four relate to the course.  This is seen in Figure \ref{fig:important:all} (c). In total there are only four predictors related to instructor in the  top-10 list of influential predictors. None of the instructor related predictors were identified in the top-5 predictors of student satisfaction in terms of the course relevancy for professional development (Q11).  This is illustrated in Figure \ref{fig:important:all} (d) where it can also be seen that only a total of four predictors related to instructor are in the list of top-10 influential predictors. Similarly, no instructor related predictor was found to influence the student satisfaction in terms of the course connecting them to real-world situations (Q12), see Figure \ref{fig:important:all}(e).





\begin{figure}[h]
\centering
\includegraphics[width=3.3in]{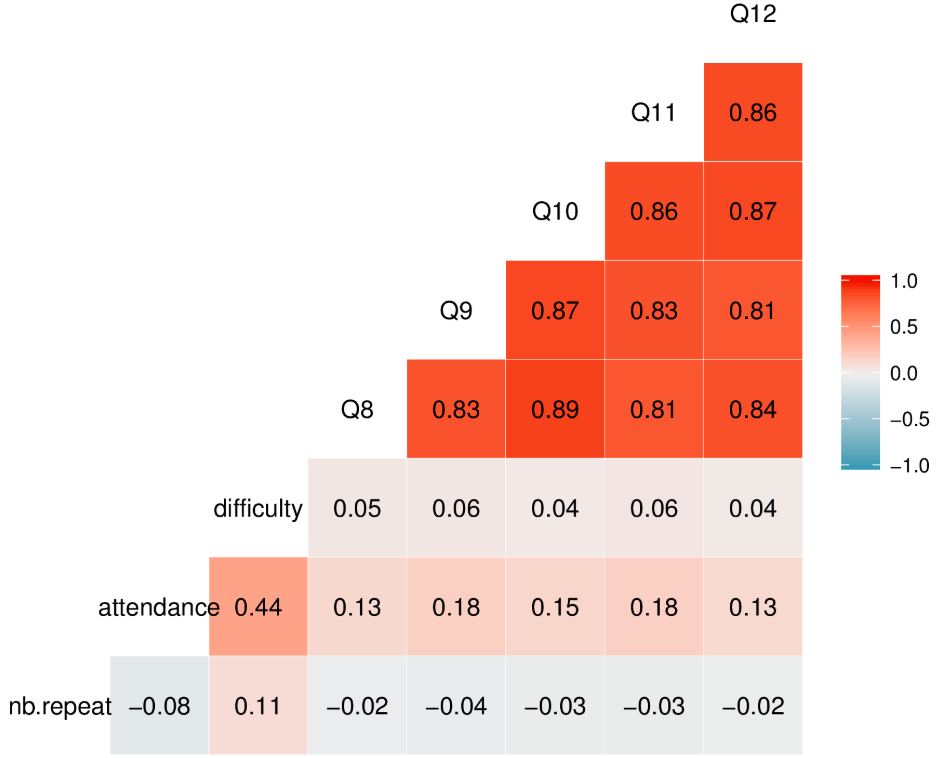}
 \caption{Correlations among different aspects of student satisfaction.}
\label{fig:corr:matrix}
\end{figure}

We further analyze the inter-correlation between the different aspects of student satisfaction using a Pearson's inter-correlation matrix. We found that there are high positive correlations between the different aspects of student satisfaction (i.e. Q8, Q9, Q10, Q11, Q12) as can be seen from the high correlation coefficients (> 0.8) illustrated in Figure \ref{fig:corr:matrix}. We also found a positive correlation (0.44) between the level of attendance in a course and its perceived level of difficulty.

\section{Conclusion}
In this paper we used machine learning techniques to infer student satisfaction from five different perspectives. We showed that the machine learning techniques deployed in our study can predict different perspectives of student satisfaction with accuracies between 80\% and 85\% when trained with both course and instructor related factors. We also investigated the top-10 influential predictors of student satisfaction. We found that the course related predictors dominate the list of top-10 as they fill the top-5 positions. This also implies that the course related factors mostly drive student satisfaction. 

This means that for this dataset, the students were satisfied in all five aspects of their course if it allowed fieldwork, applications, laboratory, discussion.  They were also happy if their class discussions, homework assignments, applications and studies were satisfactory. There is a notable influence of up to date course resources and textbooks along with the quality of class discussions, homework, and assignments. The students also prefer the course to be taught according to the syllabus announced on the first day of class with the course aims and objectives stated clearly. Only after these course considerations came class participation under committed understandable instructors who gave relevant assignments and helped, encouraged and guided the students.


%
\balance
\bibliographystyle{IEEEtran}
\bibliography{references.bib}

\end{document}